\documentclass[journal]{IEEEtran}
\ifCLASSINFOpdf
  \usepackage[pdftex]{graphicx}
  \DeclareGraphicsExtensions{.pdf,.jpeg,.png}
\else
\fi
\usepackage{tikz}
\usepackage{subfig}

\usepackage{amsmath}
\newcommand{\bld}[1]{\textbf{#1}}

\begin{document}
\title{High Resolution Time-Frequency Generation\\ with Generative Adversarial Networks}

\author{Zeynel~Deprem,
        A. Enis~Çetin,
        ~\IEEEmembership{Fellow,~IEEE}% <-this % stops a space
\thanks{Z. Deprem was with Turk Telekom, Ankara, Turkey (e-mail:zeynel.deprem@gmail.com).  He is about to leave.}% <-this % stops a space
\thanks{A.E. Çetin is with Department of Electrical and Computer Engineering, University of Illinois, Chicago, USA (e-mail: aecyy@uic.edu).}% <-this % stops a space
\thanks{Manuscript received June 1, 2021; revised June 26, 2021.}}


\markboth{Signal Processing Letters, ~Vol.~xx, No.~Y, June~2021}%
{Shell \MakeLowercase{\textit{et al.}}: Bare Demo of IEEEtran.cls for IEEE Journals}



\maketitle

\begin{abstract}
Signal representation in Time-Frequency (TF) domain is valuable in many applications including radar imaging and inverse synthetic aparture radar. TF representation allows us to identify signal components or features in a mixed time and frequency plane. There are several well-known tools, such as Wigner-Ville Distribution (WVD), Short-Time Fourier Transform (STFT) and various other variants for such a purpose. The main requirement for a TF representation tool is to give a high-resolution view of the signal such that the signal components or features are identifiable. A commonly used method is the reassignment process which reduces the cross-terms by artificially moving smoothed WVD values from their actual location to the center of the gravity for that region. In this article, we propose a novel reassignment method using the Conditional Generative Adversarial Network (CGAN). We train a CGAN to perform the reassignment process. Through examples, it is shown that the method generates high-resolution TF representations which are better than the current reassignment methods. 
\end{abstract}


% \fi
\IEEEpeerreviewmaketitle

\section{Introduction}
\IEEEPARstart{J}{oint} Time-Frequency (TF) representation of a signal reveals many features which are valuable in many applications including radar imaging \cite{Radar} and  inverse synthetic aperture radar (ISAR) \cite{TFDAndISAR3} in which a high-resolution joint-TF analysis is necessary to obtain a focused image of the target. TF representation is also valuable in many sound classification \cite{SoundClassification} and recognition \cite{TFSoundRecognition} applications. In many 
Convolutional Neural Network (CNN) sound applications \cite{CNNBasedSoundAppl}, TF representations boosts the performance of the Network.

The classical tool for the TF representation is the Wigner-Ville Distribution(WVD) \cite{WignerD}. Though it has many good features, it suffers from cross terms  which makes the readability of the TF distribution difficult and hard to interpret. Another well-known tool is the Short Time Fourier Transform (STFT) \cite{LinAndQuadTFD}. STFT does not produce any the cross terms but deteriorates the resolution of the signal. There are various other derivatives of the WVD which are named as the Cohen's Class of distributions. In addition, there are other representations which process smoothed versions of WVD by a method called as the reassignment \cite{ReassignedSpectrum1} to obtain high resolution. 

In this article a high resolution TF representation is obtained by using Generative Adversarial Networks (CGAN) \cite{goodfellow2014generative}, in particular by using a Conditional GAN (CGAN) implementation pix2pix \cite{isola2018imagetoimage} developed for general image to image transformation applications.

\section{TIME FREQUENCY REPRESENTATIONS}
The classical tool for the TF analysis is the Wigner-Ville Distribution (WVD)\cite{WignerD}. WVD of a signal $x(t)$ is given by,
\begin{equation}\label{Wigner} 
 W_x(t,f) = \int_{-\infty}^{+\infty} x(t+\tau/2)x^* (t-\tau/2) e^{-j2\pi f\tau}d\tau.
\end{equation}
WVD has many good features, such as high resolution and is a real valued distribution but due to its  quadratic nature it produces unwanted artifacts, named as cross terms, on the TF plane. This makes the readability of different signal parts on TF plane difficult.
A generalization of the WVD is the Cohen's Class \cite{CohenTFReview} distribution given by,
 \begin{equation}\label{CohenTFDEqn}
P_x(t,f) =\iint_{-\infty}^{+\infty} A_x(\theta,\tau) \Phi (\theta,\tau ) e^{-j2\pi(\theta t+f\tau)}d\theta d\tau.	
\end{equation}
where $A_x(\theta,\tau)$ is the ambiguity function (AF) of the signal $x(t)$ and has a two-dimensional (2D) Fourier transform relation with WVD. $\Phi (\theta,\tau )$ is the kernel of the distribution and has the time-frequency shaping or smoothing effect on $W_x(t,f)$. A proper kernel can attenuate the cross terms and produce high resolution distribution \cite{BaranuikOptimalKernel,MyPOCSMakale,MySIUMakale}. 

Another classical tool for the TF analysis is the Short-Time Fourier Transform (STFT) given by
\begin{equation}\label{STFT} 
 STFT_x(t,f) = \int_{-\infty}^{+\infty} x(\tau) w^*(\tau-t) e^{-j2\pi f\tau}d\tau,
\end{equation}
where, $w(t)$ is the time domain window having the smoothing effect on frequency domain. Spectrogram given by $S_x(t,f) = |STFT_x(t,f)|^2$ is a member of Cohen's Class and depending on the length of the window, can attenuate the cross terms.  But there is a trade off between the cross term suppression and the TF resolution or localization. 

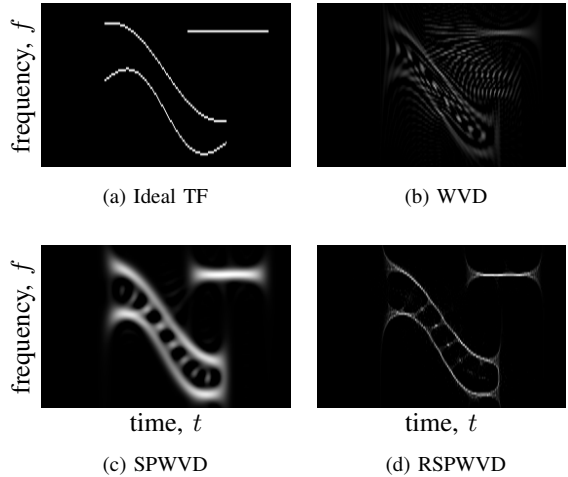
\begin{figure}[!t]
\centering
\subfloat[Ideal TF]{\begin{tikzpicture}
    \node (img1) {\includegraphics[width=1.3in]{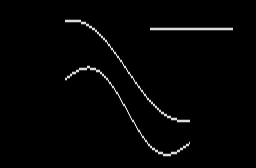}};
    \node[left of = img1, node distance=1.9cm, rotate=90, anchor=center] {frequency, $f$};
\end{tikzpicture}
\label{fig_ideal}}
\subfloat[WVD]{\begin{tikzpicture}
    \node (img2) {\includegraphics[width=1.3in]{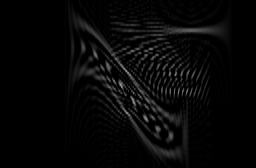}};
\end{tikzpicture}
\label{fig_WVD}}
\vfill
\subfloat[SPWVD]{\begin{tikzpicture}
    \node (img3) {\includegraphics[width=1.3in]{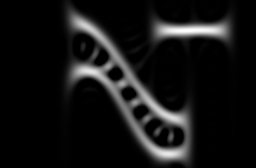}};
    \node[below of = img3, node distance = 2.8cm, yshift=1.5cm] {time, $t$};
    \node[left of = img3, node distance=1.9cm, rotate=90, anchor=center] {frequency, $f$};
\end{tikzpicture}
\label{fig_SPWVD}}
\subfloat[RSPWVD]{\begin{tikzpicture}
    \node (img4) {\includegraphics[width=1.3in]{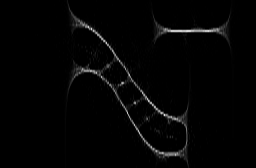}};
    \node[below of = img4, node distance = 2.8cm, yshift=1.5cm] {time, $t$};
\end{tikzpicture}
\label{fig_CGANTF}}
\caption{a) TF representation of a signal with ideal, b) WVD, c) SPWVD and d) RSPWVD, respectively.}
\label{fig_TFSample}
\end{figure}

As a remedy to the resolution problem, the reassignment method is developed \cite{ReassignedSpectrum1}. In reassignment, the value of a smoothed version, $P_x(t,f)$ of WVD  is moved to a new location $(\hat t, \hat f)$ which is the center of the gravity of the signal energy distributed around $(t,f)$. In this way, a better localization is obtained. In Figure \ref{fig_TFSample} the ideal TF, the WVD, a smoothed version of WVD, called Smoothed Pseudo WVD (SPWVD) and the reassignment of SPVWD (RSPWVD) are shown for a sample signal. As can be seen WVD is not able to reveal the three TF signal components. The SPWVD is able to show them but with a reduced resolution. But RSPWVD substantially reduces the cross terms and it also achieves high resolution or localization. Though the reassignment has high localization, it may deviate from the desired or ideal TF. In fact, at some $(t,f)$ points it may exhibit peaks which do not explain the underlying signal. This is caused by the nonlinear reassignment operation.

\section{GENERATIVE ADVERSARIAL NETWORKS}
Convolutional Neural Networks (CNN)s have revolutionized computer vision and image classification fields \cite{LecunY}.
Most  image processing applications can be regarded as a mapping from an input image to an output image. Until recent advances in  (CNN) area, the image mapping tasks have been performed by classical image processing tools \cite{ClassicalImageprocess1,eigen2015predicting}.  Convolutional autoencoders and Generative Adversarial Networks (GANs)\cite{goodfellow2014generative} have been successfully used in image generation prediction applications \cite{radford2015unsupervised}. In general, autoencoder type CNNs are trained to minimize a given cost function which determines whether an image generated by CNN is close to a desired or a given label image \cite{LecunY}. The quality of output or predicted image is heavily dependent on the selected cost function. Autoencoders trained with Euclidean ($\ell^2$) distance, or their regularized versions with the $\ell^1$ norm produce blurred predictions \cite{zhang2016colorful}. Therefore, selecting a proper cost function is crucial in CNN based image prediction problems. On the other hand, Generative Adversarial Networks (GAN) \cite{goodfellow2014generative} have a different structure with a novel cost function and they achieve sharp image reconstructions. GANs have two competing networks. The Discriminator ($D$) network tries to determine if its input is real or artificially generated by the Generator ($G$) network.  GANs are trained with a min-max optimization method given by \cite{goodfellow2014generative}
\begin{align}\label{GANcost} 
 \min_G \max_D V(D,G) &= E_{Y\sim p_Y(Y)}[\log D(Y)] \nonumber \\ 
 &+ E_{Z\sim p_Z(Z)}[\log (1 - D(G(Z)))].
\end{align}
where $Y$ represents an actual data sample (image) and $Z$ is a random vector representing the latent space whose probability distribution function (pdf) $p_Z(Z)$ is given a priori. The generator network $G$ is trained to generate an image according to its input vector $Z$. $E_{Y\sim p_{Y}(Y)}$ and $E_{Z\sim p_Z(Z)}$ are the expectation over data samples and the random variable $Z$, respectively. The network is trained in an alternating manner. In one turn $D$ is fixed and parameters of $G$ are trained, in the next round, $G$ is fixed and $D$ is trained. The aim of GAN is to estimate the pdf $p_Y(Y)$ of data so that, when fed by a random variable $Z$ the samples generated by $G$ are indistinguishable from actual data samples. In other words, the generator is trained to fool the discriminator. On the other hand, the discriminator is trained not to be fooled. The beauty of GANs come from the definition of their cost function. 
%The cost function used in GANs has a high level definition. 
Rather than measuring the distance between the desired and the generator ($G$) output with a predefined cost function the generator output is classified either as real or fake via the discriminator network $D$ . In this way, the parameters of of the generator network are "learned" during training process.  

A variant of the original GAN, which has proved success in image prediction and generation, is the Conditional GAN (CGAN). In CGANs  the generation process is based on a given condition. During training, together with the random variable $Z$, an input $X$, which define a condition on what is to be generated, is fed both to the generator and the discriminator. In this training approach, the generator is able to generate samples which are restricted by the condition $X$. In CGAN, rather than the prior distribution $p_Y(Y)$, the conditional pdf $p_{Y|X}(Y|X)$ is estimated. 
%The optimization function for CGAN is given by, 
%\begin{align}\label{cGANcost} 
 %\min_G \max_D V_C(D,G) &= E_{X,Y}[\log D(X,Y)] \nonumber \\ 
 %&+ E_{X,Z}[\log (1 - D(X,G(X,Z)))].
%\end{align}
 
\section{TF GENERATION WITH CGAN}

\begin{figure*}[!t]
\centering
\subfloat[Training with generator output (Fake) $G$]{\begin{tikzpicture}
\def\xcor{0}
\def\ycor{3}
\node[anchor=south west,inner sep=0] (G) at (\xcor,\ycor)
{\includegraphics[width=0.1\textwidth]{Figs/spwv1393.png}};
\draw[very thick] (\xcor+2.4,\ycor+0.25) rectangle node{$G$} ++(1,0.8);
\node[anchor=south west,inner sep=0] at (\xcor+4,\ycor) {\includegraphics[width=0.1\textwidth]{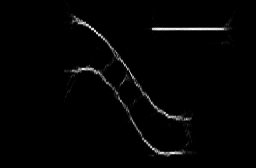}};
\node[anchor=south west,inner sep=0] at (\xcor+4,\ycor-1.4) {\includegraphics[width=0.1\textwidth]{Figs/spwv1393.png}};
\draw[very thick] (\xcor+6.4,\ycor-1) rectangle node{$D$} ++(0.8,1.7);
\draw [-latex] (\xcor+1.8,\ycor+0.7) -> ++(0.6,0);
\draw [-latex] (\xcor+3.4,\ycor+0.7) -> ++(0.6,0);
\draw [-latex] (\xcor+5.8,\ycor+0.4) -> ++(0.6,0);
\draw [-latex] (5.8,2.3) -> (6.4,2.3);
\draw [-latex] (7.2,2.8) -> (8,2.8) node[above]{Fake};
\draw [->,latex-latex](2.8,4) ++( 210 : 2.3 ) arc ( 210:300:2.3);
\node at (2.4,2.2) {$P_x(t,f)$};
\end{tikzpicture}%
\label{fig_CGAn_Fake}}
\hfil
\subfloat[Training with Ideal (Real) TF ]{\begin{tikzpicture}
\def\xcor{0}
\def\ycor{3}
\node[anchor=south west,inner sep=0] at (\xcor,\ycor) {\includegraphics[width=0.1\textwidth]{Figs/Ideal1393.png}};
\node[anchor=south west,inner sep=0] at (\xcor,\ycor-1.5) {\includegraphics[width=0.1\textwidth]{Figs/spwv1393.png}};
\draw[very thick] (\xcor+2.5,\ycor-1) rectangle node{$D$} ++(0.8,1.7);
\draw [-latex] (\xcor+1.8,\ycor+0.4) -> ++(0.7,0);
\draw [-latex] (\xcor+1.8,\ycor-0.7) -> ++(0.7,0);
\draw [-latex] (\xcor+3.3,\ycor-0.2) -> ++(0.8,0) node[above]{Real};
\node at (\xcor-0.8,\ycor+0.6) {$I_x(t,f)$};
\node at (\xcor-0.8,\ycor-1) {$P_x(t,f)$};
\end{tikzpicture}%
\label{fig_CGAN_Real}}
\caption{CGAN Training with TF distributions}
\label{fig:TFCGAN}
\end{figure*}

In this section, we consider the TF plots as two-dimensional images and generate reassigned TF plots from SPWVD distributions using CGANs.
%All the signal processing applied to original WWD to obtain a desired TF representation can be regarded as an image processing application. Therefore, the advances in image processing with GANs, CGANs in particular, can be applied to TF processing. 
Even though the aim  of CGAN is to estimate a conditional pdf $p_{Y|X}(Y|X)$, the experimental results shows that, the CGANs produce little  stochasticity at the output and the result is nearly deterministic, hence  $p_{Y|X}(Y|X)$ is impulse like \cite{928e476715544027af08ec20936dd6ca,wang2016generative}. Therefore, in some implementations the stochasticity, rather than using the random variable $Z$, is introduced in the form of "dropout" in generator implementation. In fact, pix2pix \cite{isola2018imagetoimage}, which is developed for general image to image translation applications, is implemented in this manner. In this CGAN, an image which represent the condition $X$ is fed to the generator, and the corresponding output is evaluated by the discriminator to be real or fake compared to a ground true or desired image \cite{mirza2014conditional}. Starting with this intuition, a high resolution TF generation method is proposed with CGAN as shown in Figure \ref{fig:TFCGAN}, 
where the condition $X$ or the input image to the generator, is taken as the TF distribution $P_x(t,f)$ of time domain signal $x(t)$. $P_x(t,f)$ can be any smoothed version of WVD. In this paper, the smoothed pseudo WVD (SPWVD) distribution given by 
\begin{equation}\label{SPWD}
P_x(t,f)=\iint_{-\infty}^{+\infty} h(\tau) g(s-t)R_x(s,\tau)e^{-j2\pi f\tau}ds d\tau,
\end{equation}
is used, where $R_x(s,\tau) = x(s+\tau/2)x^* (s-\tau/2)$ is the instantaneous signal auto correlation,  $g(t)$  and $h(t)$ are the time and frequency smoothing windows, respectively. SPWV is also a member of Cohen's class. $Y = G(X)$ is the generator output. As the desired or ground true image, the ideal TF trajectory $I_x(t,f)$ of the underlying signal $x(t)$ is used. The ideal TF trajectory for a synthetic signal can be easily obtained as follows. A multi-component signal which has disjoint signal components on TF plane can be constructed as
\begin{equation}\label{multicompsignal}
x(t)=\sum_{k=1}^{L} a_k (t) e^{j\phi_k(t)}
\end{equation}
where $a_k (t) e^{j\phi_k(t)}$ is the $k^{th}$ component of the signal, $a_k(t)$ is the amplitude and $\phi_k(t)$ is the phase. Though not all the signals with time-varying frequency content can be expressed in this form, most practical ones are in this form. Then the TF trajectory of the $k^{th}$ component can be expressed as,
\begin{equation}\label{TFDTrajectory_k}
I_k(t,f)=a_k^2(t) \frac{1}{2\pi} \delta \left ( f - \frac{d\phi_k (t)}{dt} \right )
\end{equation}
where $\delta()$ is the Dirac delta function and $\frac{d\phi_k (t)}{dt}$ is the instantaneous frequency (IF). We can express the ideal TF trajectory of the signal $x(t)$ as
\begin{equation}\label{TFDTrajectory}
I_x(t,f)= \sum_{k=1}^{L} I_k(t,f)
\end{equation}
Therefore, we can construct a training set for CGAN based signals whose ideal TF trajectory is known. Together with SPWVD, $P_x(t,f)$ of the signal, we can train the CGAN setup in Figure \ref{fig:TFCGAN}. After training, the aim is to generate a high resolution TF representation for an arbitrary signal. The benchmark will be the ideal TF trajectory.

The pix2pix model is a modified version of the CGAN. First, it does not use any random variable and uses the condition $X$ only as the input. The randomization is introduced as dropout in generator. Second, Pix2pix uses regularization by the $\ell^1$ norm and uses the cost function given by, \cite{isola2018imagetoimage}
\begin{align}\label{Pix2Pixcost}
 \min_G \max_D V_1(D,G) &= V_C(D,G) \nonumber \\
 &+ \lambda E_{X,Y}[\Vert Y- G(X)\Vert_1]
\end{align}
where the last term regulates the generator output with the $\ell^1$ norm. The regularization is arranged with the hyper-parameter $\lambda$. Both the generator and discriminator in pix2pix, uses the U-Net structure \cite{ronneberger2015unet}. Another variation in pix2pix implementation of CGAN is in the discriminator. The discriminator makes the real or fake decision with a PatchGAN structure where the image is divided into $N\times N$ patches. Each patch is decided whether to be real or fake and then the average of all decisions are used to give final decision about whole image.

\section{TRAINING AND TEST RESULTS}
\begin{figure}
\centering
\subfloat[Ideal TF]{\includegraphics[width=1.3in]{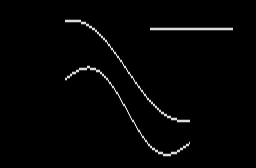}%
\label{fig_ideal1}}
\hfil
\subfloat[SPWVD]{\includegraphics[width=1.3in]{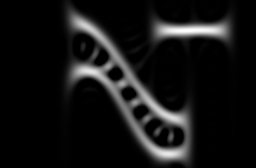}%
\label{fig_SPWV}}
\vfill
\subfloat[RSPWVD]{\includegraphics[width=1.3in]{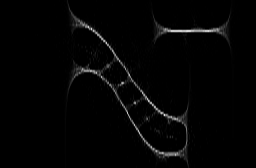}%
\label{fig_Reass}}
\hfil
\subfloat[CGAN TF]{\includegraphics[width=1.3in]{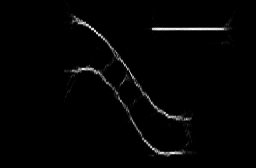}%
\label{fig_CGANTF_x1}}
\caption{Comparisson of TF distributions for a signal $x_1$ with three methods.}
\label{fig_Comp}
\end{figure}

\begin{figure}
\centering
\subfloat[Ideal TF]{\includegraphics[width=1.3in]{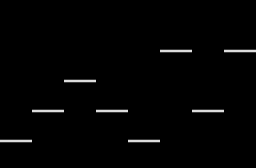}%
\label{fig2_ideal}}
\hfil
\subfloat[SPWVD]{\includegraphics[width=1.3in]{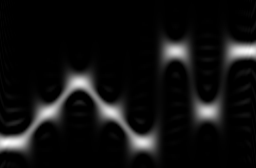}%
\label{fig2_SPWV}}
\vfill
\subfloat[RSPWVD]{\includegraphics[width=1.3in]{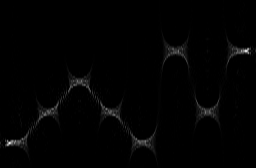}%
\label{fig2_Reass}}
\hfil
\subfloat[CGAN TF]{\includegraphics[width=1.3in]{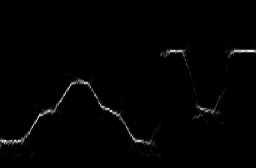}%
\label{fig2_CGANTF}}
\caption{Comparison of TF distributions for a signal $x_2$ with three methods.}
\label{fig2_Comp}
\end{figure}

\begin{figure}[!t]
\centering
\subfloat[SPWVD]{\includegraphics[width=1.4in]{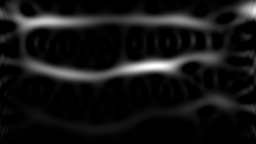}%
\label{fig2_SPWV_2}}
\vfill
\subfloat[RSPWVD]{\includegraphics[width=1.4in]{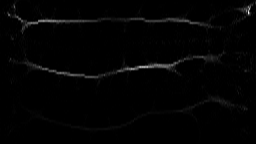}%
\label{fig3_Reass}}
\hfil
\subfloat[CGAN TF]{\includegraphics[width=1.4in]{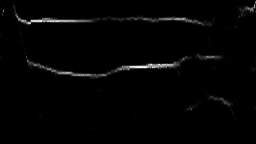}%
\label{fig3_CGANTF}}
\caption{Comparison of TF distributions for a signal obtained from Dolphin's click signal.}
\label{fig3_Comp}
\end{figure}

In order to test the effectiveness of the proposed method, a training set of 1320 signals and related SPWVD $P_x(t,f)$ and ideal TF distributions $I_x(t,f)$ were constructed using the TF toolbox \cite{TFToolBoxFlandrin}. We also used data augmentation to obtain input-output training pairs. The augmentation was done by first obtaining a smaller set and then applying various jittering operations. The time-domain signals have 256 samples. As a result the corresponding TF distributions are $256 \times 256$ images.  In addition, 120 test signals are generated  for the validation purposes. The pix2pix network \cite{isola2018imagetoimage} was implemented using
the Tensorflow Artificial Intelligence (AI) library \cite{45381}. The $\ell^1$ regularization parameter was set to $\lambda =100$ as selected by \cite{isola2018imagetoimage}. Similarly the Adam optimizer was selected with learning rate $\alpha =0.0002$, $\beta_1 =0.5$ and $\beta_2 =0.999$. The training was carried out for 50 epochs over the training set with batches of 10 images. In Figures \ref{fig_Comp}, and \ref{fig2_Comp} the result for two test signals are shown. Part (a) is the ideal TF representation of the signal $I_x(t,f)$ and (b) is the SPWVD $P_x(t,f)$.  Part (c) shows the TF obtained by reassignment of SPWVD, namely  RSPWVD. Part (d) is the proposed TF distribution obtained by pix2pix CGAN generator. Based on the Figures \ref{fig_Comp} and \ref{fig2_Comp} it is clear that  pix2pix CGAN produces high resolution TFs whose performance is comparable to the ideal TF. In order to asses the performance of the proposed CGAN based method a quantitative comparison is also carried out and the results are shown in Table \ref{Table1}. 
\begin{table*}[!t]
\renewcommand{\arraystretch}{1.3}
\caption{Performance comparison for SPVWD, RSPWVD and CGAN methods.}
\label{Table1}
\centering
\begin{tabular}{|c|ccc|ccc|ccc|ccc|ccc|}
\hline
 & \multicolumn{3}{|c|}{\textbf{Ideal TF}}& \multicolumn{3}{|c|}{\textbf{WVD}}& \multicolumn{3}{|c|}{\textbf{SPWVD}}& \multicolumn{3}{|c|}{\textbf{RSPWVD}} & \multicolumn{3}{|c|}{\textbf{CGAN TF}}\\
\hline
& pc $\uparrow$ & $\ell^1 \downarrow$ & R $\downarrow$ & pc $\uparrow$& $\ell^1 \downarrow$  & R $\downarrow$ & pc $\uparrow$ & $\ell^1 \downarrow$  & R $\downarrow$ & pc $\uparrow$ & $\ell^1 \downarrow$  & R $\downarrow$ & pc $\uparrow$ & $\ell^1 \downarrow$  & R $\downarrow$ \\
\hline
$x_1$ & 1 & 0 & 9.82 & 0.27 & 1735 & 12.79 & 0.43 & 3606 & 12.88 & \bld{0.52} & 820 & 10.80 & 0.49 & \bld{753} & \bld{10.32}\\
$x_2$ & 1 & 0 & 9.64 & 0.46 & 816 & 10.82 & 0.57 & 1948 & 12.07 & 0.60 & 424 & 9.76 & \bld{0.84} & \bld{265} & \bld{9.89}\\
$x_3$ & 1 & 0 & 9.71 & 0.36 & 1067 & 11.66 & 0.56 & 1975 & 12.12 & 0.54 & 536 & 10.06 & \bld{0.78} & \bld{328} & \bld{9.85}\\
$x_4$ & 1 & 0 & 8.64 & 0.29 & 1036 & 12.52 & 0.34 & 1823 & 12.22 & \bld{0.57} & 392 & 9.97 & 0.34 & \bld{361} & \bld{9.34}\\
$x_5$ & 1 & 0 & 11.02 & 0.27 & 3757 & 13.87 & 0.49 & 5039 & 13.51 & 0.59 & 1524 & 11.37 & \bld{0.67} & \bld{1325} & \bld{11.29}\\
$x_6$ & 1 & 0 & 9.49 & 0.20 & 2052 & 12.80 & 0.44 & 2667& 12.40& 0.42 & 719 & 10.71 & \bld{0.49} & \bld{611} & \bld{9.78}\\
$x_7$ & 1 & 0 & 9.87 & 0.30 & 1570 & 12.58 & 0.48 & 3056 & 12.64 & 0.47 & 924 & 10.72 & \bld{0.73} & \bld{535} & \bld{10.09}\\
$x_8$ & 1 & 0 & 10.04 & 0.41 & 2379 & 12.80 & 0.55 & 3225 & 12.47 & 0.52 & 846 & \bld{10.29} & \bld{0.71} & \bld{656} & 10.44\\
$x_9$ & 1 & 0 & 10.07 & 0.39 & 1927 & 13.06 & 0.45 & 4537 & 13.05 & \bld{0.61} & 849 & 10.65 & 0.59 & \bld{869} & \bld{10.55}\\
$x_{10}$ & 1 & 0 & 10.10 & 0.31 & 2381 & 13.21 & 0.52 & 3352 & 12.70 & 0.64 & 797 & 10.57 & \bld{0.74} & \bld{615} & \bld{10.37}\\
\hline
$\ldots$ & $\ldots$ & $\ldots$ & $\ldots$ & $\ldots$ & $\ldots$ & $\ldots$ & $\ldots$ & $\ldots$ & $\ldots$ & $\ldots$ & $\ldots$ & $\ldots$ & $\ldots$ & $\ldots$ & $\ldots$\\
\hline
$Avg(120)$ & 1 & 0 & 9.82 & 0.33 & 1852 & 12.60 & 0.48 & 3075 & 12.58 & 0.55 & 771 & 10.48 & \bld{0.64} & \bld{610} & \bld{10.16}\\
\hline
\end{tabular}
\end{table*}
The first 10 rows ($x_1, x_2, \hdots, x_{10}$) shows the result for 10 test signals. The bottom line is  obtained by averaging the results for 120 test signals. Three metrics are considered. The first metric is the Pearson correlation (pc) \cite{PearsonCorr} coefficient which is given by,
\begin{equation}\label{Pearson}
  pc = \frac{ P_x^T I_x }{\Vert P_x \Vert_2 \Vert I_x \Vert_2}
\end{equation}
where $P_x$ and $I_x$ are the vector form of related discrete TF matrices with the mean subtracted. Pearson correlation measures the shape similarity between the vectors. The higher the Pearson correlation the better the similarity. The arrow next to each metric indicates the desired direction. The second metric is the $\ell^1$ difference between the method and the ideal TF. The third measure is the Renyi entropy \cite{RenyiEntropy} which measures the TF localization. Renyi entropy is given by 
\begin{equation}\label{Renyi}
    R_{P}^\alpha = \frac{1}{1-\alpha}\log_2{\sum_{n=0}^N\sum_{m=0}^M P_x^\alpha[m,n]}.
\end{equation}
where $\alpha$ is the order of Renyi. A Renyi entropy of order 3 was shown to be a good measure for localization \cite{RenyiEntropy}. The lower the Renyi measure the better the localization. 
Table I clearly shows that the results obtained with the proposed CGAN approach are better than all the existing reassignment methods. For each metric the method performing the best is is printed in bold. Even though, the localization of RSPWVD and CGAN based method are comparable, CGAN based method is better in terms of $\ell^1$ measure and the Pearson correlation. This is attributed to the fact RSPWVD may have some $(t,f)$ points which do not express the underlying signal due to the nonlinear reassignment operation. Yet another comparison is made for a signal obtained from a Dolphin’s click-signal segment. The result is shown in Figure \ref{fig3_Comp}. Since it is not a synthetic signal the ideal TF is not known a priori, hence, is not shown. This figure also shows that the method successfully generates the TF representation.
\section{Conclusion}

A method which generates high resolution TF representations was obtained using CGAN. The method uses SPWVD of the signal. Through examples it was shown  that the performance is better than the existing methods, in particular the method gives better results than the reassignment method.

% or


%\section*{Acknowledgment}


\ifCLASSOPTIONcaptionsoff
  \newpage
\fi


%\bibliographystyle{IEEEtran}

\bibliographystyle{IEEEtran}
\bibliography{IEEEabrv,refs}

%\newpage


%\vfill


\end{document}